\documentclass[sigconf,screen]{acmart}
\usepackage{xurl}
\usepackage{tabularx}
\AtBeginDocument{%
  }

\newcommand{\datasetname}{CAS2UML}

\setcopyright{cc}
\setcctype{by}
\acmDOI{10.1145/3832783.3834613}
\acmYear{2026}
\copyrightyear{2026}
\acmISBN{979-8-4007-2882-2/2026/10}
\acmConference[ASE '26]{Proceedings of the 41st IEEE/ACM International Conference on Automated Software Engineering}{October 12--16, 2026}{Munich, Germany}
\acmBooktitle{Proceedings of the 41st IEEE/ACM International Conference on Automated Software Engineering (ASE '26), October 12--16, 2026, Munich, Germany}
\acmSubmissionID{ase26tool-p54-p}
\received{2026-05-12}
\received[accepted]{2026-06-19}

\begin{document}

\title{CAS2UML: A Handwritten Sketch-to-PlantUML Dataset for Class and Activity Diagrams}

\author{Simon Scholz}
\orcid{0009-0005-2663-2377}
\affiliation{%
  \institution{University of Cologne}
  \city{Cologne}
  \country{Germany}
}
\email{sschol24@smail.uni-koeln.de}

\author{Mersedeh Sadeghi}
\correspondingauthor
\orcid{0000-0001-6405-8824}
\affiliation{%
  \institution{University of Cologne}
  \city{Cologne}
  \country{Germany}
}
\email{sadeghi@cs.uni-koeln.de}

\renewcommand{\shortauthors}{Scholz et al.}


\begin{abstract}
Automated UML generation from sketches and images is gaining renewed attention with the rise of large language models and multimodal AI. However, reproducible evaluation remains difficult due to the lack of public datasets with executable ground-truth models. We present \datasetname{}, a public dataset of 557 hand-drawn UML diagrams, including 271 class diagrams and 286 activity diagrams, each paired with manually validated PlantUML code. We also provide a PlantUML-based validation tool and reusable scripts for checking the syntactic correctness and renderability of generated UML artifacts, enabling reproducible benchmarking of sketch-to-UML approaches.
The dataset, validation tool, processing scripts, documentation, and demonstration video are publicly available at: 
\textbf{Dataset:} \url{https://huggingface.co/datasets/Seym0n/cas2uml_hand-drawn_to_plantuml_dataset}; 
\textbf{Tool and Scripts:} \url{https://github.com/Seym0n/handwritten-uml-dataset}; 
\textbf{Video:} \url{https://www.youtube.com/watch?v=KQrYeGgT3hs}.
\end{abstract}

\ccsdesc[500]{Software and its engineering~Unified Modeling Language (UML)}
\ccsdesc[500]{Computing methodologies~Machine learning}

\keywords{dataset, UML, handwritten UML, PlantUML, benchmark, sketch recognition, software modeling, large language models}

\maketitle

\section{Introduction}
The Unified Modeling Language (UML) remains a central notation for software specification, analysis, design, and communication among stakeholders~\cite{ozkaya2020survey,mohagheghi2009definitions}. Among its diagram types, class and activity diagrams are particularly important: class diagrams are consistently reported as the most frequently used UML notation, while activity diagrams are among the most common behavioral notations used for workflow and process modeling~\cite{koc2021uml,Reggio2014}. Together, they cover complementary structural and behavioral views of software systems.

The automatic creation of UML models from informal inputs has therefore been a long-standing goal in software engineering. Earlier work explored model generation from textual requirements, rendered diagrams, and scanned documentation, while recent advances in multimodal AI and large language models have renewed interest in generating UML from prompts, screenshots, photographs, and hand-drawn sketches~\cite{zhao2021nlp4re,karasneh2013extracting,chen2023automated,de2024evaluating,conrardy2024image2uml}. These developments suggest that sketch-to-UML and image-to-model pipelines are becoming increasingly feasible.

However, progress is limited by the lack of public benchmark datasets that pair hand-drawn UML diagrams with executable, machine-readable ground truth. Existing resources are either image-only, restricted to visual annotations, limited to small in-house benchmarks, or focused on a single diagram type~\cite{piucco2024,conrardy2024image2uml}. This makes it difficult to evaluate whether generated UML artifacts are not only visually plausible, but also syntactically valid, editable, and reusable in modeling toolchains.

To address this gap, we introduce \datasetname{}, a public dataset of 557 hand-drawn UML
diagrams, consisting of 271 class diagrams and 286 activity diagrams. Each sample is paired
with validated PlantUML\footnote{https://plantuml.com/} code, providing an executable textual
representation of the corresponding sketch. PlantUML was selected because it is human-readable,
editable, renderable, and already used as an output format in recent UML-generation
studies~\cite{conrardy2024image2uml,alahmad2025student}.

In addition, we provide a validation tool and scripts that support the creation, rendering, and
syntactic checking of sketch--PlantUML pairs. Together, the dataset and tool enable reproducible
evaluation of sketch-recognition, image-to-UML, multimodal AI, and automated model-generation
approaches.

In summary, this paper contributes (i) the \datasetname{} dataset of validated
sketch--PlantUML pairs, and (ii) a validation tool and reusable scripts for checking PlantUML
renderability, syntactic correctness, and dataset integrity.

\section{Background and Related Works}
\begin{table*}[t]
\centering
\footnotesize
\setlength{\tabcolsep}{4pt}
\caption{Comparison of UML sketch datasets.}
\label{tab:dataset_comparison}
\begin{tabularx}{\textwidth}{@{}l l l r X c c@{}}
\toprule
\textbf{Dataset / Study} & \textbf{Source} & \textbf{Diagram type(s)} &
\textbf{\#\,samples} & \textbf{Ground truth} & \textbf{Validated} & \textbf{Public} \\
\midrule
Conrardy \& Cabot~\cite{conrardy2024image2uml} & Hand-drawn & UML class & 4 & PlantUML & Partial & No \\
Piucco~\cite{piucco2024} & Hand-drawn & UML class & 70 & Element masks & -- & Yes \\
Ra\v{z}inskas et al.~\cite{ravzinskas2024transforming} & Hand/tool-drawn & UML use case & 346 & Bounding boxes & -- & Yes \\
\midrule
\textbf{Ours} & \textbf{Hand-drawn} & \textbf{Class + activity} & \textbf{557} & \textbf{PlantUML (+XMI for class)} & \textbf{Yes} & \textbf{Yes} \\
\bottomrule
\end{tabularx}
\end{table*}
\subsection{Class and Activity Diagrams}

Class diagrams and activity diagrams were selected because they represent two of the most widely used UML notations. Class diagrams capture the static structure of a system, including classes, attributes, operations, and relationships, and are often used for domain modeling, architectural design, and implementation-oriented documentation~\cite{karasneh2013extracting,chen2023automated}. They are consistently reported as the most frequently studied and most broadly covered UML artifact across the research literature, books, tools, and courses~\cite{koc2021uml,Reggio2014}.

Activity diagrams complement this structural view by modeling behavior, workflows, decisions, parallel flows, and process logic. Their broad set of control-flow constructs makes them widely used for business-process modeling, requirements analysis, and service-oriented and workflow-based systems~\cite{Reggio2014}. By covering both class and activity diagrams, \datasetname{} supports evaluation across structural and behavioral UML modeling tasks.

\subsection{Limitations of Existing Datasets}
Despite growing interest in generating UML models from sketches and diagram images, reproducible evaluation remains a major challenge. Existing studies on image-based or sketch-based UML creation rely on very small evaluation sets. Related LLM-based UML studies also typically evaluate on small, task-specific benchmarks, such as ten domain descriptions in Chen et al.~\cite{chen2023automated} and twenty class-diagram exercises in De Bari et al.~\cite{de2024evaluating}.
Also Eklund and Jonsson~\cite{eklund2025benchmarking} benchmark LLMs on UML generation from informal notations, but evaluate model behaviour rather than releasing paired sketch–ground-truth data; CAS2UML is complementary in providing the resource such benchmarks require.

Table~\ref{tab:dataset_comparison} focuses on resources that are closest to our setting: hand-drawn UML sketches. Large mined repositories such as Lindholmen~\cite{hebig2016umlgithub} contain many UML images, but they primarily represent rendered or collected diagram images rather than hand-drawn sketches. They therefore address an adjacent image-based UML setting rather than the sketch-to-UML setting targeted in this paper. Existing hand-drawn UML resources are much smaller and are usually limited to a single diagram type. Conrardy and Cabot~\cite{conrardy2024image2uml} used only four hand-drawn class-diagram examples, while Piucco~\cite{piucco2024} provides 70 handwritten class-diagram images with visual annotations, but no PlantUML or XMI ground truth. Ra\v{z}inskas et al.~\cite{ravzinskas2024transforming} provide a public collection of UML use-case sketches with element-level annotations; however, the released dataset is not paired with executable UML ground truth for each sketch. In contrast, \datasetname{} provides 557 hand-drawn diagrams across class and activity diagrams, each paired with validated executable PlantUML code.

In summary, three observations emerge from the literature review. First, in neighboring modeling domains such as BPMN, public hand-drawn datasets already support evaluation at the scale of several hundred diagrams~\cite{schafer2021sketch2bpmn, schafer2023sketch2process}, whereas comparable UML resources remain limited in size and often provide only visual annotations. Second, recent UML-generation studies typically rely on small, task-specific, and often unreleased evaluation sets. Third, no prior dataset combines hand-drawn UML inputs with editable, machine-readable, and end-to-end validated ground truth at a scale suitable for systematic benchmarking. \datasetname{} is designed to fill this gap by enabling sketch-recognition, sketch-to-UML, and automated model-generation approaches to be evaluated against the same executable reference.

\section{\datasetname{} Data Creation}
The dataset was constructed to provide paired examples of hand-drawn UML diagrams and executable PlantUML specifications. The creation workflow followed four main steps: (i) defining diagram-content specifications, (ii) generating or collecting initial PlantUML representations, (iii) creating handwritten diagram images from validated references, and (iv) verifying the one-to-one correspondence between each sketch and its machine-readable ground truth. The first author performed all manual steps.

\subsection{Class Diagram}
\begin{table*}[h]
\centering
\caption{Class-diagram coverage guideline: UML constructs and GoF-inspired patterns used during sample creation. Numbers in parentheses indicate how often a pattern was redrawn in a distinct context.}
\label{tab:class_diagram_guideline}
\small
\begin{tabularx}{\linewidth}{@{}p{0.23\linewidth}X@{}}
\toprule
\textbf{Category} & \textbf{Covered elements / patterns (Instances)} \\
\midrule
\textit{Creational patterns} &
Singleton (3), Factory Method (3), Abstract Factory (2), Builder (1), Prototype (1) \\
\addlinespace[0.10em]
\textit{Structural patterns} &
Adapter (3), Bridge (2), Composite (2), Decorator (3), Facade (3), Flyweight (1), Proxy (2) \\
\addlinespace[0.10em]
\textit{Behavioral patterns} &
Observer (2), Strategy (3), Command (2), State (3), Template Method (2), Chain of Responsibility (2), Iterator (2), Mediator (2), Memento (2), Visitor (2) \\
\addlinespace[0.10em]
\textit{Association types} &
Unidirectional association, bidirectional association, self-association, ternary association \\
\addlinespace[0.10em]
\textit{Aggregation and composition} &
Simple aggregation, composition, mixed aggregation/composition \\
\addlinespace[0.10em]
\textit{Inheritance hierarchies} &
Single inheritance with 2--3 levels, multiple inheritance, interface implementation, abstract classes with concrete subclasses, mixed inheritance and composition \\
\addlinespace[0.10em]
\textit{Cardinality variations} &
One-to-one, one-to-many, many-to-many, optional, bounded ranges\\
\addlinespace[0.10em]
\textit{Dependencies} &
Dependency relation using dashed arrows \\
\bottomrule
\end{tabularx}
\end{table*}
The class-diagram subset was created through a two-source process. First, we reused the 70 handwritten class-diagram images released by Piucco~\cite{piucco2024}. Since this dataset provides visual annotations but no executable UML representation, we created corresponding PlantUML code for these samples and validated the resulting sketch--PlantUML pairs using the procedure described in Section~\ref{subsec:dataset-processing}.

Second, we created 208 additional class-diagram samples to increase the size, diversity, and construct coverage of the dataset. For this purpose, we defined a coverage guideline specifying the UML elements and modeling situations to be represented. The guideline includes common class-diagram constructs such as associations, aggregation, composition, inheritance, interfaces, dependencies, multiplicities, and abstract classes. It also incorporates examples inspired by the Gang of Four design patterns in order to obtain realistic object-oriented structures with varying complexity~\cite{hunt2013gang}. Table~\ref{tab:class_diagram_guideline} summarizes the resulting coverage guideline.

\subsection{Dataset Processing, Validation, and Release}
\label{subsec:dataset-processing}

For each additional class-diagram sample, we first created a reference PlantUML specification using an LLM-assisted authoring workflow with ChatGPT 5.3, Claude Sonnet 4.6 and Gemini 3 Flash. The generated code was manually inspected, corrected where necessary, and rendered using PlantText\footnote{https://www.planttext.com/} to verify that the resulting diagram matched the intended class-diagram structure. The validated reference diagrams were then manually redrawn on different media, including tablets, plain paper, and checkered paper. This process produced hand-drawn images with variation in writing style, line quality, background, and acquisition conditions.

Each final sample, therefore, consists of a handwritten class-diagram image paired with validated executable PlantUML code. This process resulted in 278 class-diagram pairs before XMI filtering and 271 released class-diagram samples after excluding seven diagrams whose PlantUML code could not be serialized to XMI due to unsupported association-class constructs.
\begin{figure}[t]
    \centering
    \includegraphics[width=0.85\linewidth]{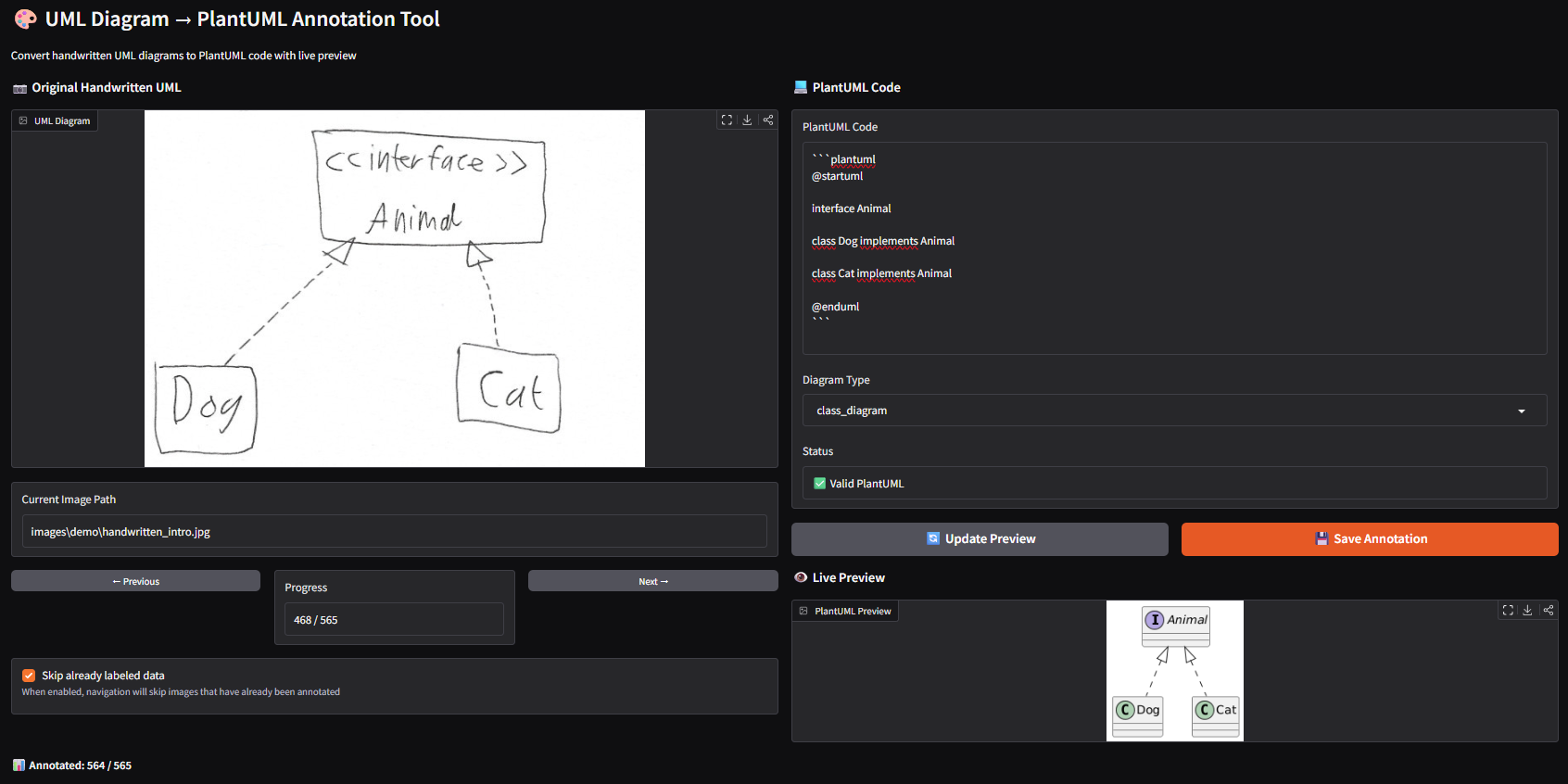}
    \caption{The \datasetname{} validation tool with side-by-side sketch input, PlantUML annotation, and rendered preview.}
    \label{fig:data-annotation-tool-screenshot}
\end{figure}

\subsection{Activity Diagram}
For the activity-diagram subset, we constructed the data from scratch, since no comparable public hand-drawn activity-diagram dataset was available. The goal was to obtain a diverse set of activity diagrams that covers common control-flow constructs while remaining suitable for manual redrawing and end-to-end validation.

We used two sources for creating reference diagrams. First, we considered the Lindholmen dataset~\cite{hebig2016umlgithub}, a large collection of UML diagram images mined from GitHub repositories. To identify candidate activity diagrams, we filtered the dataset metadata using URL-path keywords, in particular occurrences of ``activity'' together with common image-file extensions. This filtering step yielded approximately 1,000 candidate activity-diagram images. From these candidates, around 80 diagrams were selected as reference templates based on readability, relevance, and suitability for redrawing.

Second, to increase construct coverage beyond the selected repository mined examples, we created additional activity-diagram specifications covering common UML activity constructs, including actions, initial and final nodes, decision and merge nodes, fork and join nodes, loops, and swimlanes. PlantUML code for these diagrams was produced through an LLM-assisted authoring workflow using Claude Sonnet 4.6 and Gemini 3 Flash, and rendered using PlantText. The generated diagrams were then manually inspected and corrected to ensure that the resulting activity flows were syntactically valid and semantically coherent.

The final handwritten images were created by manually redrawing the validated reference diagrams on different media, including tablets, blank paper, and checkered paper. This process introduced variation in handwriting style, layout, line quality, background, and acquisition conditions. Each handwritten activity diagram was then paired with its corresponding validated PlantUML specification.

In total, the activity-diagram subset contains 286 paired samples, each consisting of a handwritten activity-diagram image and an executable PlantUML representation.

\subsection{\datasetname{} Validator Tool}
To support the creation and verification of the sketch--PlantUML pairs, we developed a custom web-based validation tool. The tool displays each handwritten UML image next to an editable PlantUML panel and a rendered preview of the corresponding diagram. This setup allows one to iteratively edit the PlantUML code, render it locally using PlantUML, and check whether the executable representation matches the handwritten input. Existing annotation tools did not provide the required combination of sketch visualization, editable PlantUML, and live UML rendering; therefore, we implemented a lightweight interface using Gradio.

Figure~\ref{fig:data-annotation-tool-screenshot} shows the validation interface. For each sample, the tool records the image path, diagram type, and validated PlantUML code. The diagram type is specified as either class diagram or activity diagram. After saving an annotation, the tool stores the pair in a JSON file and allows navigation through the remaining samples. This workflow ensures that each released sample contains both a handwritten image and an executable PlantUML representation.

After annotation, the dataset was processed for public release. All image paths were verified against the annotation file, non-image files were excluded, and all images were resized so that the longest side did not exceed 1000 pixels. This step reduced variation caused by different acquisition conditions, such as scanned paper drawings and tablet-generated images, while preserving visual readability.

To ensure that the released PlantUML annotations are executable, we applied an additional validation procedure. Each annotation was checked using the PlantUML JAR with the \texttt{-syntax} flag to verify that the code is parseable and free of compilation errors. For class diagrams, we also generated StarUML-compatible\footnote{https://staruml.io/} XMI files from the validated PlantUML code using the PlantUML server. This step was applied only to class diagrams, since PlantUML does not support XMI export for activity diagrams. 
To document the visual diversity of the released dataset, we also report the drawing media used for the handwritten samples in Table~\ref{tab:medium_distribution}. The dataset includes tablet drawings, plain-paper sketches, checkered-paper sketches, photographed paper diagrams, and one lined-paper sample. This variation supports evaluation under different visual conditions and acquisition settings.

\begin{table}[h]
    \centering
    \caption{Distribution of drawing media in \datasetname{}.}
    \label{tab:medium_distribution}
    \begin{tabular}{lrrr}
        \toprule
        \textbf{Medium} & \textbf{Class} & \textbf{Activity} & \textbf{Total} \\
        \midrule
        Tablet & 149 & 120 & 269 \\
        Plain paper & 43 & 52 & 95 \\
        Checkered paper & 8 & 114 & 122 \\
        Photographed paper$^*$ & 70 & 0 & 70 \\
        Lined paper & 1 & 0 & 1 \\
        \midrule
        \textbf{Total} & \textbf{271} & \textbf{286} & \textbf{557} \\
        \bottomrule
        \multicolumn{4}{l}{\footnotesize $^*$Sourced from Piucco~\cite{piucco2024}.}
    \end{tabular}
\end{table}

\begin{figure}[h]
    \centering
    \setlength{\abovecaptionskip}{4pt}
    \setlength{\belowcaptionskip}{0pt}
    \includegraphics[
        page=1,
        trim=2mm 2mm 111mm 42mm,
        clip,
        width=0.8\columnwidth
    ]{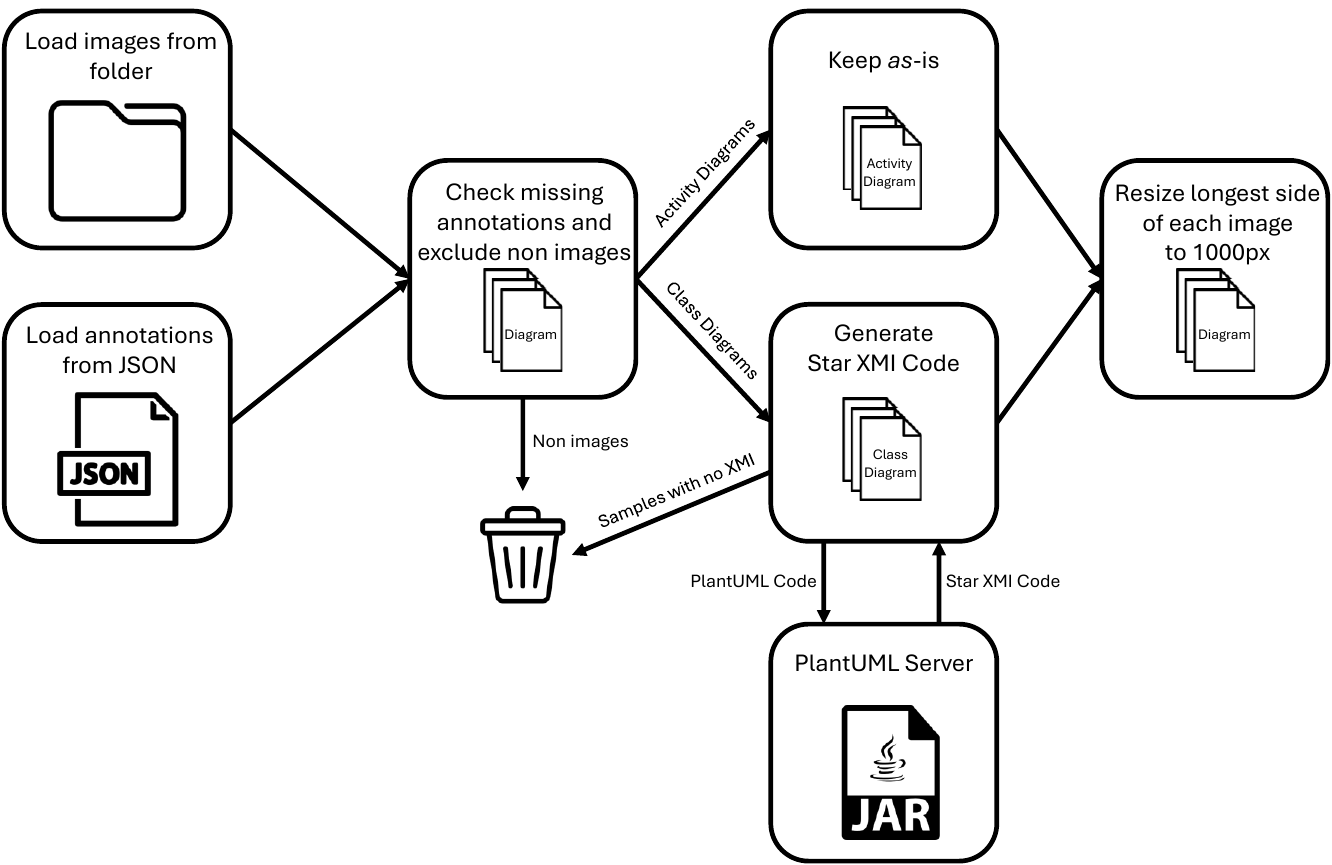}
    \caption{\datasetname{} processing and validation pipeline}
    \label{fig:process_hf}
\end{figure}

Finally, the processed images, PlantUML code, diagram-type labels, and available XMI files were uploaded to the HuggingFace Hub\footnote{https://huggingface.co/}. We also release the validation script, which produces machine-readable error reports and can be reused to check PlantUML outputs generated by sketch-to-UML systems. The overall processing and release workflow is summarized in Figure~\ref{fig:process_hf}.

As preliminary evidence of utility, we use CAS2UML in follow-up work to fine-tune a handwritten-UML-to-PlantUML model that, evaluated via automatic metrics and a human ranking study, becomes competitive with proprietary vision-language baselines~\cite{sadeghi2026formalism}.

\section{Conclusion}

This paper introduced \datasetname{}, a public dataset of 557 hand-drawn class and activity UML diagrams paired with validated executable PlantUML code, together with a validation tool and processing pipeline for creating and checking sketch--PlantUML pairs. By providing an executable ground-truth reference, \datasetname{} enables reproducible benchmarking of sketch-recognition, image-to-UML, and automated model-generation approaches, and supports future research on AI-assisted software modeling.

\bibliographystyle{ACM-Reference-Format}
\bibliography{citations} 

\end{document}